\documentclass[12pt]{article}
\usepackage{amsfonts,amssymb,latexsym}
\usepackage{indentfirst}
\usepackage{amscd,amsopn, amsmath }
 \usepackage{feynmp}
 \usepackage[pdftex]{graphicx}
\newcommand{\LL}{ {\cal L} }

\newcommand{\sh}[1]{#1\hskip-7pt \diagup}
\newcommand{\Sh}[1]{#1\hskip-11pt \diagup}
\newcommand{\ssh}[1]{#1\hskip-9pt \diagup}

\newcommand{\eq}{\begin{equation}}
\newcommand{\eqn}{\end{equation}}

\newcommand{\tr}{{\rm Tr}}

\newcommand{\half}{\frac{1}{2}}
\newcommand{\halfi}{\frac{i}{2}}
\newcommand{\sen}{{\rm sin}}

\begin{document}
\begin{fmffile}{diagras}
\fmfcmd{%
     style_def wiggly_arrow expr p=
	cdraw (wiggly p);
	shrink (1.5);
	  cfill (arrow p);
	endshrink;
     enddef;}

%%%%%%%%%%%%%

\newcommand{\propagluon}{
	{\begin{fmfgraph*}(45,25)
                \fmfleft{i}\fmfright{o}
                \fmf{wiggly_arrow, label=$k$}{i,o}
		\fmflabel{$\mu$}{i}
		\fmflabel{$\nu$}{o}
%                \fmflabel{{\footnotesize y}}{o}
%                \fmflabel{{\footnotesize x}}{i}
                \end{fmfgraph*}}}

\newcommand{\propalambda}{
	{\begin{fmfgraph*}(45,25)
                \fmfleft{i}\fmfright{o}
                \fmf{plain_arrow, label=$k$}{i,o}
		%\fmflabel{$\mu$}{i}
		%\fmflabel{$\nu$}{o}
%                \fmflabel{{\footnotesize y}}{o}
%                \fmflabel{{\footnotesize x}}{i}
                \end{fmfgraph*}}}

\newcommand{\propachi}{
	{\begin{fmfgraph*}(45,25)
                \fmfleft{i}\fmfright{o}
                \fmf{scalar, label=$k$}{i,o}
		%\fmflabel{$\mu$}{i}
		%\fmflabel{$\nu$}{o}
%                \fmflabel{{\footnotesize y}}{o}
%                \fmflabel{{\footnotesize x}}{i}
                \end{fmfgraph*}}}

\newcommand{\propaghost}{
	{\begin{fmfgraph*}(45,25)
                \fmfleft{i}\fmfright{o}
                \fmf{ghost, label=$k$}{i,o}
		%\fmflabel{$\mu$}{i}
		%\fmflabel{$\nu$}{o}
%                \fmflabel{{\footnotesize y}}{o}
%                \fmflabel{{\footnotesize x}}{i}
                \end{fmfgraph*}}}

\newcommand{\propaboson}{
	{\begin{fmfgraph*}(45,25)
                \fmfleft{i}\fmfright{o}
                \fmf{dbl_dots, label=$k$}{i,o}
		\fmflabel{$\mu$}{i}
		\fmflabel{$\nu$}{o}
%                \fmflabel{{\footnotesize y}}{o}
%                \fmflabel{{\footnotesize x}}{i}
                \end{fmfgraph*}}}

\newcommand{\tresgvertex}{
	{\begin{fmfgraph*}(55,55)
                \fmfleft{i}\fmfright{o1,o2}
                \fmf{wiggly_arrow,label=$p$}{i,v1}
		\fmf{wiggly_arrow,label=$q$}{o1,v1}
		\fmf{wiggly_arrow,label=$r$}{o2,v1}
                \fmflabel{$k$, $\mu$}{i}
                \fmflabel{$l$, $\nu$}{o1}
		\fmflabel{$m$, $\rho$}{o2}
                \end{fmfgraph*}}}

\newcommand{\fourgvertex}{
	{\begin{fmfgraph*}(55,55)
                \fmfleft{i1,i2}\fmfright{o1,o2}
                \fmf{wiggly_arrow,label=$q$}{i1,v1}
		\fmf{wiggly_arrow,label=$p$}{i2,v1}
		\fmf{wiggly_arrow,label=$r$}{o1,v1}
		\fmf{wiggly_arrow,label=$s$}{o2,v1}
                \fmflabel{$k$, $\mu$}{i2}
                \fmflabel{$l$, $\nu$}{i1}
		\fmflabel{$m$, $\rho$}{o1}
		\fmflabel{$n$, $\delta$}{o2}
                \end{fmfgraph*}}}

\newcommand{\lamgvertex}{
	{\begin{fmfgraph*}(55,55)
                \fmfleft{i}\fmfright{o1,o2}
                \fmf{wiggly_arrow,label=$p$}{i,v1}
		\fmf{fermion,label=$q$}{o1,v1}
		\fmf{fermion,label=$r$}{v1,o2}
                \fmflabel{$k$, $\mu$}{i}
                \fmflabel{$l$,$a$}{o1}
		\fmflabel{$m$,$b$}{o2}
                \end{fmfgraph*}}}

\newcommand{\chigvertex}{
	{\begin{fmfgraph*}(55,55)
                \fmfleft{i}\fmfright{o1,o2}
                \fmf{wiggly_arrow,label=$p$}{i,v1}
		\fmf{scalar,label=$q$}{o1,v1}
		\fmf{scalar,label=$r$}{v1,o2}
                \fmflabel{$k$, $\mu$}{i}
                \fmflabel{$l$}{o1}
		\fmflabel{$m$}{o2}
                \end{fmfgraph*}}}

\newcommand{\ghgvertex}{
	{\begin{fmfgraph*}(55,55)
                \fmfleft{i}\fmfright{o1,o2}
                \fmf{wiggly_arrow,label=$p$}{i,v1}
		\fmf{ghost,label=$q$}{o1,v1}
		\fmf{ghost,label=$r$}{v1,o2}
                \fmflabel{$k$, $\mu$}{i}
                \fmflabel{$l$}{o1}
		\fmflabel{$m$}{o2}
                \end{fmfgraph*}}}

\newcommand{\tsgvertex}{
	{\begin{fmfgraph*}(55,55)
                \fmfleft{i}\fmfright{o1,o2}
                \fmf{wiggly_arrow,label=$p$}{i,v1}
		\fmf{dbl_dots,label=$q$}{o1,v1}
		\fmf{dbl_dots,label=$r$}{o2,v1}
                \fmflabel{$k$, $\mu$}{i}
                \fmflabel{$l$,$a$}{o1}
		\fmflabel{$m$,$b$}{o2}
                \end{fmfgraph*}}}

\newcommand{\fsgvertex}{
	{\begin{fmfgraph*}(55,55)
                \fmfleft{i1,i2}\fmfright{o1,o2}
                \fmf{wiggly_arrow,label=$q$}{i1,v1}
		\fmf{wiggly_arrow,label=$p$}{i2,v1}
		\fmf{dbl_dots,label=$r$}{o1,v1}
		\fmf{dbl_dots,label=$s$}{o2,v1}
                \fmflabel{$k$, $\mu$}{i2}
                \fmflabel{$l$, $\nu$}{i1}
		\fmflabel{$m$,$a$}{o1}
		\fmflabel{$n$,$b$}{o2}
                \end{fmfgraph*}}}

\newcommand{\slamvertex}{
	{\begin{fmfgraph*}(55,55)
                \fmfleft{i}\fmfright{o1,o2}
                \fmf{dbl_dots,label=$p$}{i,v1}
		\fmf{fermion,label=$q$}{o1,v1}
		\fmf{fermion,label=$r$}{v1,o2}
                \fmflabel{$k$,$a$}{i}
                \fmflabel{$l$,$b$}{o1}
		\fmflabel{$m$,$c$}{o2}
                \end{fmfgraph*}}}

\newcommand{\slamchivertex}{
	{\begin{fmfgraph*}(55,55)
                \fmfleft{i}\fmfright{o1,o2}
                \fmf{dbl_dots,label=$p$}{i,v1}
		\fmf{scalar,label=$q$}{o1,v1}
		\fmf{fermion,label=$r$}{v1,o2}
                \fmflabel{$k$,$a$}{i}
                \fmflabel{$l$}{o1}
		\fmflabel{$m$,$b$}{o2}
                \end{fmfgraph*}}}

\newcommand{\tscalvertex}{
	{\begin{fmfgraph*}(55,55)
                \fmfleft{i}\fmfright{o1,o2}
                \fmf{dbl_dots,label=$p$}{i,v1}
		\fmf{dbl_dots,label=$q$}{o1,v1}
		\fmf{dbl_dots,label=$r$}{o2,v1}
                \fmflabel{$k$,$a$}{i}
                \fmflabel{$l$,$b$}{o1}
		\fmflabel{$m$,$c$}{o2}
                \end{fmfgraph*}}}

\newcommand{\fscalvertex}{
	{\begin{fmfgraph*}(55,55)
                \fmfleft{i1,i2}\fmfright{o1,o2}
                \fmf{dbl_dots,label=$q$}{i1,v1}
		\fmf{dbl_dots,label=$p$}{i2,v1}
		\fmf{dbl_dots,label=$r$}{o1,v1}
		\fmf{dbl_dots,label=$s$}{o2,v1}
                \fmflabel{$k$,$a$}{i2}
                \fmflabel{$l$,$b$}{i1}
		\fmflabel{$m$,$c$}{o1}
		\fmflabel{$n$,$d$}{o2}
                \end{fmfgraph*}}}

\newcommand{\loopuno}{
	{\begin{fmfgraph*}(120,25)
                \fmfleft{i}\fmfright{o}
                \fmf{wiggly}{i,v1}
	 	\fmf{wiggly}{v2,o}
		\fmf{wiggly,left}{v1,v2,v1}
                \end{fmfgraph*}}}

\newcommand{\loopdos}{
	{\begin{fmfgraph*}(120,25)
                \fmfleft{i}\fmfright{o}
                \fmf{wiggly}{i,v1}
	 	\fmf{wiggly}{v2,o}
		\fmf{plain,left}{v1,v2,v1}
                \end{fmfgraph*}}}

\newcommand{\looptres}{
	{\begin{fmfgraph*}(120,25)
                \fmfleft{i}\fmfright{o}
                \fmf{wiggly}{i,v1}
	 	\fmf{wiggly}{v2,o}
		\fmf{dashes,left}{v1,v2,v1}
                \end{fmfgraph*}}}

\newcommand{\loopfour}{
	{\begin{fmfgraph*}(120,25)
                \fmfleft{i}\fmfright{o}
                \fmf{wiggly}{i,v1}
	 	\fmf{wiggly}{v2,o}
		\fmf{dots,left}{v1,v2,v1}
                \end{fmfgraph*}}}

\newcommand{\loopfive}{
	{\begin{fmfgraph*}(120,25)
                \fmfleft{i}\fmfright{o}
                \fmf{wiggly}{i,v1}
	 	\fmf{wiggly}{v2,o}
		\fmf{dbl_dots,left}{v1,v2,v1}
                \end{fmfgraph*}}}

\newcommand{\loopsix}{
	{\begin{fmfgraph*}(120,25)
                \fmfleft{i}\fmfright{o}
                \fmf{wiggly}{i,v,v,o}
	 	%\fmf{wiggly}{v2,o}
		%\fmf{,left}{v1,v2,v1}
                \end{fmfgraph*}}}

\newcommand{\loopseven}{
	{\begin{fmfgraph*}(120,25)
                \fmfleft{i}\fmfright{o}
                \fmf{wiggly}{i,v1}
	 	\fmf{wiggly}{v1,o}
		\fmf{dbl_dots}{v1,v1}
                \end{fmfgraph*}}}

\vspace{.5cm}
\begin{center}
\Large{\bf One-loop Renormalized Coefficient of Noncommutative Supersymmetric Yang-Mills-Chern-Simons Gauge Theories in Three Dimensions}\\
\vspace{.5cm}

\large Sendic Estrada-Jim\'enez$^{a}$\footnote{e-mail: {\tt
sestrada@unach.mx}}, Hugo
Garc\'{\i}a-Compe\'an$^{b}$\footnote{e-mail: {\tt
compean@fis.cinvestav.mx}}\\
[4mm]

{\small \em $^a$ Centro de Estudios  en F\'{\i}sica y  Matem\'aticas B\'asicas y Aplicadas, \\
 Universidad Aut\'onoma de Chiapas,
Calle 4$^a$ Oiente Norte 1428, Tuxtla Guti\'errez, Chiapas,
M\'exico}
\\
[4mm]
{\small \em $^b$ Departamento de F\'{\i}sica, Centro de Investigaci\'on y de Estudios Avanzados del IPN}\\
{\small\em P.O. Box 14-740, 07000, M\'exico D.F., M\'exico}\\[4mm]

\vspace*{.5cm}
\small{\bf Abstract} \\
\end{center}

\begin{center}
\begin{minipage}[h]{14.0cm} {
Recent studies of the $AdS_4/CFT_3$
correspondence involve the construction of a peculiar
supersymmetric gauge theory on the worldvolume of multiple M2s branes
as a boundary field theory. Under suitable conditions the quantum theory
becomes a noncommutative supersymmetric YM-CS gauge theory which call for an study
of its renormalized perturbative corrections. As a preliminary step to more general
consideration, the modification of the ${\cal N}=3,2,1$ supersymmetric YM-CS gauge theory
due to noncommutativity of spatial coordinates is proposed. We carry out the one-loop renormalization and a noncommutative correction for the Chern-Simons coefficient is obtained. Finally it is found that this new correction depends of the noncommutative parameter in an analytic form. }
\end{minipage}
\end{center}

%\bigskip
%\bigskip

\date{\today}
%\leftline{CINVESTAV-FIS/02-027}
%\leftline{\tt hep-th/yymmnnn}

%\vspace{1cm}

\leftline{June 21, 2011}

\newpage

\section{Introduction}

In recent years, supersymmetric Chern-Simons (CS) gauge theories  have attracted a great deal of attention due to the correspondence ${\rm AdS}_4/{\rm CFT}_3$ between CS matter field theories (CSM) and M-theory on $AdS_4\times S^7$ (the ABJM model  \cite{Maldacena}). It is expected that a superconformal CSM theory with a large number of supersymmetries be useful  to describe, at low energies, the worldvolume theory on multiple membranes (M2-branes) in M-theory.  However in Ref. \cite{swartz2004}, it was argued that these theories have not the required supersymmetries. Moreover from the construction of a model with ${\cal N}=8$ supersymmetry \cite{bagger1} (the BL model) a lot of work has been developed in different directions (for instance see
\cite{Copland:2010yx} and references therein).  On the other hand, it has been constructed a large class of
${\cal N}=  4$ CSM theories by a method that enhances ${\cal N} =1$
supersymmetry to ${\cal N} = 4$ \cite{GW}, and has been proved that with some suitable conditions these theories are equivalent to the model building in \cite{bagger1}.  By using group representation theory, from  ${\cal N }= 1$ to ${\cal N} = 8$ CSM theories were constructed systematically \cite{elena}, and the equivalence of these models has been described for ${\cal N}=5$ in \cite{wu2010}.

In Ref. \cite{bedfor} it is studied the quantum properties of the theory of Bagger and Lambert (BL) where it is analyzed the perturbative shift in the  CS coupling constant. They use a Yang-Mills action as regulator in the spirit of \cite{kao}, and find that there are a one-loop correction in the coupling $\kappa
\to \kappa+2sgn(\kappa)$. They conjecture that, although the BL theory and the model proposed in \cite{Maldacena} for ${\cal N}=6$ are equivalent classically, they may not be equivalent at the quantum level. Another  study in the context of quantum properties of CSM models for ${\cal N}=2$ is performed in \cite{bianchi}. So the quantum properties of CS theories with supersymmetry are interesting.

Perturbative studies of Chern-Simons theories have many motivations. Historically they arise from the quest of new topological invariants order by order in perturbation theory \cite{semenof}.
From a seminal paper \cite{deser}, it is a known fact that the requirement of invariance of the Chern-Simons Lagrangian under finite gauge transformation leads to the quantization of the coupling constant. This quantization is also valid in the noncommutative version as it was shown in \cite{nair,wu,jabbari}. Nevertheless if one couples Yang-Mills theory in three dimensions with the Chern-Simons theory it was recognized that a shift of the coupling constant is found due to quantum corrections, $\kappa \to \kappa + c_v$, where $c_v$ is the Casimir of the underlying group. This shift is found through the analysis of the renormalization of the coupled theory \cite{semenof,pisarski}.

Supersymmetric YM-CS theories arise also from some configurations of $D3$-branes and $(p,q)$-fivebranes in Type IIB superstring theory. These theories has been described in \cite{ohta} for which is placed a $D3$-brane between NS5-branes and $D5$ branes. In \cite{ohta2,hanany} it was constructed the brane configuration which describe supersymmetric YM-CS  and the conditions under which is breaking the supersymmetry. There were reproduced the results obtained by Witten by computing the index \cite{Witten:1999ds}.

For ${\cal N}=3,2,1$ supersymmetric theories the quantum corrections for YMCS theory are nice computed in by Kao, Lee and Lee in Ref. \cite{kao}. They found a shift in the coupling constant only for ${\cal N}=1$. In the present paper we construct a noncommutatve version of Kao, Lee and Lee model. We find some no-trivial correction to the Chern-Simons coefficient in terms of the non-commutative parameter $\Theta$, which is an analytical function of this parameter. This would be relevant in order to find a noncommutative version of the the $AdS_4/CFT_3$. The field theory version would involves a noncommutative YM-CS theory of the form considered in this paper or in general grounds a noncommutative version of the BL model or the ABJM model. Some recent proposals in this direction are found in \cite{ncmembranas}. To construct the noncommutative theory we will consider only spatial noncommutativity to avoid causality problems \cite{problemas}. The noncommutativity is introduced as usual, by through the Moyal star product (for a review, see \cite{szabo,douglas}). As it is known the noncommutativity changes the algebra of the gauge group to the universal enveloping algebra of the group. As we will shown this change can be summarize in a new $\Theta$ dependent  functions of structure.

In the context of  noncommutative supersymmetric Chern-Simons theories, recently there have been some studies shown the consistency an finiteness of this kind of theories by using superfields formulation \cite{asano,ferrari1,ferrari2,ferrari3,ferrari4}.

The paper is organized as follows. I Section II we review the  supersymmetric YM-CS theory and build the noncommutative version. Section III is devoted to study the Ward-Slanov-Taylor identities. In section IV we analyze the one-loop renormalization  of our model. In section V we compute the noncommutative shift to the Chern-Simons coefficient. In section VI the final comments are presented.

%%%%%%%%%%%%%%%%%%%%%%%%%%%%%%%
%%%%%%%%%%%%%%%%%%%%%%%%%%%%%%%
%%%%%%%%%%%%%%%%%%%%%%%%%%%%%%%
\section{Noncommutative Supersymmetric Yang-Mills-Chern-Simons}

We start from the ${\cal N}=3$ supersymmetric YM Lagrangian with gauge group $G$ \cite{kao} with a explicit symmetry $O(3)$. In this Lagrangian we have the gauge multiplet, consisting of a massive vector $A_\mu$,  three Majorana Fermions $\lambda_a$, three neutral scalar bosons $C_a$ and one Majorana fermion of opposite helicity $\chi$. This Lagrangian can be obtained from the dimensional reduction for a pure supersymmetric ${\cal N}=2$ YM theory in four dimensions  \cite{kao2}. The Lagrangian is given by
\begin{eqnarray}
{\cal L}_{YM}=\frac{1}{g^2}\tr \left\{-\half F_{\mu\nu}F^{\mu\nu}+ D_\mu C_a D^\mu C_a + (D_a)^2 +
i\bar{\lambda}_a \Sh D_\mu \lambda_a +\bar{\chi}\Sh D_\mu \chi \right. \nonumber \\
+ \left. i\varepsilon_{abc}\bar{\lambda}_a[\lambda_b,C_c]-2i\bar{\lambda}_a[\chi,C_a]-
\half[C_a,C_b][C_b, C_a]\right\},
\end{eqnarray}
where $D_\mu=\partial_\mu-i[A_\mu, \cdot ]$, $a,b,c=1,2,3$ y $D_a$
are auxiliary fields. The auxiliary fields are absent when we consider
the Lagrangian on-shell. The generators of the gauge group satisfy
$[T^m,T^n]=if^{lmn}T^l$ and $\tr T^mT^n=\delta^{mn}/2$ with $f^{lmn}$ being the structure constants of
$G$. The fields
belong to the adjoint representation and $A_\mu=A_\mu^m T^m$. The
quadratic Casimir $c_f$ of the gauge group $G$ in the adjoint representation is given by
$f^{kmn}f^{lmn}=c_f\delta^{kl}$. The metric is written as
 $(1,-1,-1)$ and $\varepsilon^{012}=
\varepsilon_{012}=1$.
The gamma matrices are purely imaginary and satisfy the relation: $\gamma^\mu\gamma^\nu
= \eta^{\mu\nu}-i\varepsilon^{\mu\nu\rho}\gamma_\rho.$

Now the ${\cal N}=3$ supersymmetric Chern-Simons Lagrangian which is obtained  and  given in \cite{kao2}
\eq {\cal L}_{CS}=
\kappa \tr \left\{\varepsilon^{\mu\nu\rho}\left(A_\mu\partial_\nu
A_\rho- \frac{2}{3}iA_\mu A_\nu A_\rho
\right)-\bar{\lambda}_a\lambda_a + \bar{\chi}\chi +2C_aD_a+
\frac{i}{3} \varepsilon_{abc}C_a[C_b,C_c]\right\}, \eqn
where $\kappa$ is the coupling constant also termed the Chern-Simons coefficient.

The system to be considered in this paper comes from the addition of both Lagrangians
\begin{equation}
{\cal L} = {\cal L}_{YM} + {\cal L}_{CS}.
\end{equation}
The ${\cal N}=3$ supersymmetric transformations are given by
\begin{equation}\begin{array}{rcl}
\delta A_\mu&=& -i\bar{\alpha}_a\gamma_\mu\lambda_a, \\
\delta \lambda_a & = &i\Sh B  \alpha_a - \varepsilon_{abc} (D_b \alpha_c
-i \Sh D C_b \alpha_c ) + i[C_a,C_b]\alpha_b ,\\
\delta \chi & =& -i\Sh D C_a \alpha_a - D_a \alpha_a + \halfi \varepsilon_{abc}
[C_b,C_c] \alpha_a ,\\
\delta C_a & = &- \varepsilon_{abc} \bar{\alpha}_b\lambda_c +
\bar{\alpha}_a\chi ,\\
\delta D_a & =& i \varepsilon_{abc} \bar{\alpha}_b \Sh D \lambda_c + i
\bar{\alpha}_a \Sh D \chi +i[\bar{\alpha}_b \lambda_a, C_b] \\
& &\,\,
-i[\bar{\alpha}_b \lambda_b, C_a] +i [\bar{\alpha}_a \lambda_b, C_b]
- i\varepsilon_{abc}\bar{\alpha}_b[\chi,C_c],
\end{array}
\end{equation}
where $B^\mu=\varepsilon^{\mu\nu\rho}\partial_\nu A_\rho$.

Using field equations for the auxiliary field $D_a+\kappa g^2C_a=0,$ derived from ${\cal L}$ we can eliminate the auxiliary fields $D_a$ to obtain in this way, the total on-shell Lagrangian reads
\begin{eqnarray}
{\cal L}&=& \frac{1}{g^2}\tr \left\{-\half F_{\mu\nu}F^{\mu\nu}+ (D_\mu C_a)^2 +
i\bar{\lambda}_a\gamma^\mu D_\mu\lambda_a +i\bar{\chi}\gamma^\mu D_\mu\chi \right. \nonumber\\
& & \qquad \quad+ \left. i\varepsilon_{abc}\bar{\lambda}_a[\lambda_b, C_c]-
2i\bar{\lambda}_a[\chi, C_a]- \half [C_a,C_b][C_b,C_a] \right\} \nonumber \\
&& + \,\kappa \tr \left\{ \varepsilon^{\mu\nu\rho} \left(A_\mu \partial_\nu A_\rho-
\frac{2}{3}iA_\mu A_\nu A_\rho \right) -\kappa g^2 C_a^2 -\bar{\lambda}_a\lambda_a +
\bar{\chi}\chi \right. \nonumber \\
& &  \qquad \quad\, -\left. \frac{i}{3}\varepsilon_{abc}C_a[C_b,C_c]\right\}.
\end{eqnarray}
If we scale the gauge field by  $A_\mu^m\rightarrow
gA_\mu^m$, we can see that the expansion parameter  is $g^2$, which
has mass dimension.

We must add the fixing gauge term and the Faddeev-Popov one for the ghost fields
\eq
{\cal L}_{gf}=-\frac{1}{2\xi}(\partial^\mu A_\mu^m)^2,
\eqn
\eq {\cal L}_{FP}=-2\tr[\bar{\eta}(\partial^\mu D_\mu)\eta]. \eqn
These terms complete the commutative theory.

We are interested in analyzing the one-loop corrections of the noncommutative theory.  The spatial noncommutativity of space
is introduced by changing the usual product of smooth functions by the Moyal star product. After defining
$m=kg^2$ and adding all Lagrangians we have
\begin{eqnarray}
{\cal L}&=&\tr \frac{1}{g^2}\left\{\half \left(\partial_\mu A_\nu -
\partial_\nu A_\mu -i[A_\mu,A_\nu]_\star \right)\left(\partial^\mu A^\nu -
\partial^\nu A^\mu -i[A^\mu,A^\nu]_\star \right) \right.\nonumber\\
&&+ (\partial_\mu C_a-i[A_\mu,C_a]_\star)(\partial^\mu C_a-i[A^\mu,C_a]_\star) \nonumber \\
&& + i\bar{\lambda}_a\gamma^\mu\partial_\mu\lambda_a +
\bar{\lambda}_a\gamma^\mu[A_\mu,\lambda_a]_\star+ i\bar{\chi}\gamma^\mu\partial_\mu\chi +
\bar{\chi}\gamma^\mu[A_\mu,\chi]_\star \nonumber \\
&& + i\varepsilon_{abc}\bar{\lambda}_a[\lambda_b,C_c]_\star- 2i\bar{\lambda}_a[\chi,C_a]_\star -
\half[C_a,C_b]_\star[C_b,C_a]_\star \nonumber \\
&& + m\varepsilon^{\mu\nu\rho}\left(A_\mu \partial_\nu A_\rho -
\frac{i}{3}A_\mu[A_\nu,A_\rho]_\star \right)-m^2C_a^2-m\bar{\lambda}_a\lambda_a+
m \bar{\chi}\chi \nonumber \\
&&\left. -\frac{i}{3}\varepsilon_{abc} C_a[C_b,C_c]_\star \right\}-
\frac{1}{g^2}\frac{1}{\xi}\partial^\mu A_\mu^m\partial^\nu A_\nu^m - \bar{\eta}^m
\partial^\mu\partial_\mu\eta^m- i\partial^\mu\bar{\eta}^m[A_\mu,\eta]^m_\star,
\end{eqnarray}
where we have  omitted explicitly one star product according to the properties of it \cite{szabo,douglas}. We must remark that this is a noncommutative non-abelian theory. It is well known that when the noncommutativity is introduced in an abelian theory, the effect is,
to turns out the commutative theory into non-abelian one, with gauge symmetry being described by a universal enveloping algebra of the gauge Lie algebra \cite{wu,ferrari1,caporaso,slanov,bonora1,armoni,buch,glenn,grosse}.

Now it is necessary to see how the commutator algebra changes for the noncommutative gauge
theories. We know that the star commutator of two fields is \eq
[A_\mu,A_\nu]_\star=A_\mu\star A_\nu- A_\nu\star A_\mu, \eqn as we are working in the adjoint representation
$A_\mu=A_\mu^mT^m$, with the explicit calculus we have
\begin{eqnarray}
[A_\mu,A_\nu]_\star&=&A_\mu^mT^m\star A_\nu^n T^n- A_\nu^n T^n \star A_\mu^mT^m \nonumber \\
&=& A_\mu^m\star A_\nu^n \half ([T^m,T^n]+\{T^m,T^n\})-
A_\nu^n  A_\mu^m\half ([T^n,T^m]+\{T^n,T^m\})\nonumber \\
&=& \half A_\mu^m e^{\halfi \overleftarrow{\partial}\Theta \overrightarrow{\partial}}
A_\nu^n ([T^m,T^n]+\{T^m,T^n\}) \nonumber \\
&& - \half A_\mu^m e^{-\halfi \overleftarrow{\partial}\Theta \overrightarrow{\partial}}
A_\nu^n([T^n,T^m]+\{T^n,T^m\}).
\label{conmu}
\end{eqnarray}
Recall that the structure constants  totally
antisymmetric  $f^{klm}$ and  the  totally symmetric  $d^{klm}$ of the gauge group
$G=U(N)$ are given by the next relations \cite{masoud,compean}
\begin{equation}
[T^l,T^m]=if^{klm}T^k, \qquad \qquad \{T^l,T^m\}=d^{klm}T^k,
\end{equation}
and we can rewrite (\ref{conmu}) as
\begin{equation}
[A_\mu,A_\nu]_\star = iA^m_\mu\cos\left(\frac{\overleftarrow{\partial_\alpha}\Theta^{\alpha \beta}
\overrightarrow{\partial_\beta}}{2} \right) A_\nu^n \, f_{lmn}T^l +i A^m_\mu
\sin\left(\frac{\overleftarrow{\partial_\alpha}\Theta^{\alpha \beta}\overrightarrow{\partial_\beta}}{2}
\right) A_\nu^n \,d_{lmn}T^l.
\end{equation}
In the momentum space the last expression takes the form
\begin{equation}
[A_\mu,A_\nu]_\star = \int_{p,q} iA^m_\mu(p)\left[\cos\left(-\frac{p\wedge
q}{2} \right)\, f_{lmn}T^l +
\sin\left(-\frac{p \wedge q}{2}
\right)  \,d_{lmn}T^l\right] A_\nu^n(q) e^{i(p+q)x},
\end{equation}
where $p\wedge q \equiv p_\alpha \Theta^{\alpha \beta}q_\beta$.
Thus we can define a new structure functions as follows
\begin{equation}
F_{lmn}(q\wedge p) = f_{lmn}\cos\left(\frac{q\wedge p}{2}\right) +
d_{lmn}\sin \left(\frac{q\wedge p}{2}\right).
\end{equation}
Then we can write the commutator in a simplified form by
\eq [A_\mu,A_\nu]_\star^m= \int_{p,q}A^k_\mu(p)
A_\nu^l(q) iF_{klm}(q\wedge p)e^{-i(p+q)x}, \eqn where as we
mentioned earlier,  we are working
with the universal enveloping algebra of the gauge group
\cite{chai,wess1, wess2, wess3, wess4,jackiw2,szabo2}. The new structure
function have the following property \eq F_{lmn}(p\wedge q)=
-F_{mln}(q\wedge p). \eqn
Consequently the free Lagrangian is given by
\begin{eqnarray}
{\cal L}_0 &=& \frac{1}{2g^2}A^{m\mu}\left\{(\partial^2 \eta_{\mu\nu}-\partial_\mu\partial_\nu)-
m\varepsilon_{\mu\nu\rho}\partial^\rho +\frac{1}{\xi}\partial_\mu\partial_\nu\right\}A^{m\nu} \nonumber \\
&& + \frac{1}{2g^2}C_a(-\partial^2-m^2)C_a +\frac{1}{2g^2}\bar{\lambda}_a(i\ssh \partial - m)\lambda_a +
\frac{1}{2g^2}\bar{\chi}(i\ssh \partial +m)\chi \nonumber \\
&&+\bar{\eta}^m(-\partial^2)\eta^m,
\end{eqnarray}
while the interacting Lagrangian is written as
\begin{eqnarray}
\LL_I &=& \frac{1}{g^2}\int_k \bigg\{
i k_{1\mu} A_\nu^m(k_1) A^{n\mu}(k_2) A^{t\nu(k_3)} F_{nrm}(k_2\wedge k_3)
e^{-i(k_1+k_2+k_3)x}   \nonumber \\
&&-\frac{1}{4} A_\mu^n(k_1)A_\nu^r(k_2)A^{s\mu}(k_3)A^{t\nu}(k_4)F_{nrm}(k_1\wedge k_2)
F_{stm}(k_3\wedge k_4) e^{-i(k_1+k_2+k_3+k_4)x} \nonumber \\
&&+ \frac{m}{6}\varepsilon^{\mu\nu\rho}A_\mu^m(k_1)A_\nu^n(k_2)A_\rho^r(k_3)F_{nrm}(k_2\wedge k_3)
e^{-i(k_1+k_2+k_3)x}\nonumber \\
&&-i k_{1\mu}C_a^m(k_1)A^{n\mu}(k_2)C_a^r(k_3)F_{nrm}(k_2\wedge k_3) e^{-i(k_1+k_2+k_3)x} \nonumber \\
&&+\half A_\mu^n(k_1)C_a^r(k_2)F_{nrm}(k_1\wedge k_2) A^{s\mu}(k_3)C_a^t(k_4)F_{stm}(k_3\wedge k_4)
e^{-i(k_1+k_2+k_3+k_4)x} \nonumber \\
&&+  \frac{i}{2} \bar{\lambda}_a^m(k_1)\gamma^\mu A_\mu^n(k_2)\lambda_a^r(k_3)
F_{nrm}(k_2\wedge k_3) e^{-i(k_2+k_3-k_1)x}   \nonumber \\
&& + \halfi \bar{\chi}^m(k_1)\gamma^\mu A_\mu^n(k_2) \chi^r(k_3)F_{nrm}(k_2\wedge k_3)
e^{-i(k_2+k_3-k_1)x} \nonumber \\
&&-\half \varepsilon_{abc}\bar{\lambda}_a^m(k_1)\lambda_b^n(k_2)C_c^r(k_3)F_{nrm}(k_2\wedge k_3)
e^{-i(k_2+k_3-k_1)x} \nonumber \\
&&+ \bar{\lambda}_a^m(k_1)\chi^n(k_2)C_a^r(k_3)F_{nrm}(k_2\wedge k_3) e^{-i(k_2+k_3-k_1)x} \nonumber \\
&&+\frac{1}{4} C_a^n(k_1)C_b^r(k_2)C_b^s(k_3)C_a^t(k_4)(k_4)F_{nrm}(k_1\wedge k_2)
F_{stm}(k_3\wedge k_4) e^{-i(k_1+k_2+k_3+k_4)x} \nonumber \\
&&+\frac{1}{6}m\varepsilon_{abc}C_a^m(k_1)C_b^n(k_2)C_c^r(k_3)F_{nrm}(k_2\wedge k_3)
e^{-i(k_1+k_2+k_3)x} \bigg\}\nonumber \\
&+& \int_k i k_1^\mu \bar{\eta}^m(k_1)A_\mu^n(k_2)\eta^r(k_3) F_{nrm}(k_2\wedge k_3) e^{-i(k_2+k_3-k_1)x}.
\end{eqnarray}
Now we are in position to calculate the Feynman rules for the
theory (see Appendix A). Let us
write here only the propagator for the gauge field by
\eq
\Delta_{\mu\nu}(k)=\frac{g^2}{k^2(k^2-m^2)}(k_\mu k_\nu-k^2\eta_{\mu\nu}-
im\varepsilon_{\mu\nu\rho}k^\rho)+g^2\xi\frac{k_\mu k_\nu}{k^4}.
\eqn
In order to avoid infrared divergences we will take the Landau gauge i.e.  $\xi=0$.

%%%%%%%%%%%%%%%%%%%%%%%%%%%%%%%
%%%%%%%%%%%%%%%%%%%%%%%%%%%%%%%
%%%%%%%%%%%%%%%%%%%%%%%%%%%%%%%
\section{The Ward-Slanov-Taylor identities}

In the ordinary gauge field theory the Ward-Slanov-Taylor identities
play a very important role in the renormalizability of the perturbative
theory. For renormalizable gauge theories these identities
essentially represent the manifestation of the gauge invariance with the
regularized or renormalized action with counterterms included.
Conversely, by verifying the Ward-Slanov-Taylor identities we can check
the renormalizability and the gauge invariance of the renormalized
theory. The same is valid for the noncommutative theories \cite{wu}.
In this section we comment on the conditions that Ward identities must be fulfilled in order to verify the
gauge invariance.

Due to the symmetry of the system, we can factorize the self-energy as
\eq
\Pi_{\mu\nu}(p)=\frac{1}{m}(\delta_{\mu\nu}p^2-p_\mu p_\nu)\Pi_e-
i\varepsilon_{\mu\nu\rho}p^\rho \Pi_0.
\label{selfgl}
\eqn
Contracting  $\Pi_{\mu\nu}$ with  $\frac{\delta_{\mu\nu}}{2p^2}$
and $\frac{\varepsilon_{\mu\nu\rho}p^\rho}{2p^2}$ we obtain $\Pi_e$ and $\Pi_o$ respectively.
The kinetic term in the effective action for the gauge boson leads to
\eq
\Delta_{\mu\nu}^{-1}(p)=\Delta_{0\mu\nu}^{-1}(p)+\Pi_{\mu\nu}(p),
\eqn
where $\Pi_{\mu\nu}$ is the self-energy of the gauge boson, and the subindex $_0$
stands for the bare propagator. In the same way, the ghost propagator is
corrected in the next form
\eq
\widetilde{\Delta}(p)=\frac{1}{\widetilde{Z}(p)p^2},
\eqn
where
 \eq
\widetilde{Z}(p)=1+\widetilde{\Pi}(p).
\eqn

The part of the action that is similar to the classical Lagrangian
can be written in terms of the renormalized fields and their
respective parameters according to the standard normalization
\cite{zinn1,zinn2}. Thus we obtain the relation between the
renormalized fields and bare fields, for instance
\begin{eqnarray}
A_\mu^m&=&\sqrt{Z_3}A_{ren\,\mu}^m,\\
\eta^m&=&\sqrt{\widetilde{Z}}\eta_{ren}^m.
\end{eqnarray}
Consequently the interaction between the ghost fields and the gauge field must be the
identity after the renormalization by the Ward identities, then we
have
\eq
Z_3=\widetilde{Z}^{-2}.
\eqn

Let us define now $Z_\kappa\equiv 1- \Pi_o(p)/\kappa$ \cite{kao},
and the renormalized Chern-Simons coefficient is
\begin{eqnarray}
\kappa_{ren}&=&\kappa Z_\kappa Z_3 =Z_\kappa \widetilde{Z}^{-2} \nonumber \\
&=&\kappa \left(1-\frac{1}{\kappa}\Pi_o(p)+2\widetilde{\Pi}(p)\right).
\end{eqnarray}

%%%%%%%%%%%%%%%%%%%%%%%%%%%%%%%%%%%%%
%%%%%%%%%%%%%%%%%%%%%%%%%%%%%%%%%%%%%
%%%%%%%%%%%%%%%%%%%%%%%%%%%%%%%%%%%%%
\section{One-loop renormalization}

In here we calculate the one-loop self-energy of the
gauge field, for which there are seven diagrams, but according to
decomposition did in  (\ref{selfgl}), we have that for the odd part
$\Pi_o$ of the self-energy only contribute three diagrams,
which are those that have a term with a factor $\varepsilon^{\mu\nu\rho}$. For the
even part of the self-energy $\Pi_e$ it is necessary to take into account all diagrams.

As we seen in the previous section, to calculate the correction
to the Chern-Simons coefficient  only is necessary to find the
odd part of the self-energy of the gluon and the self-energy of the
ghost field. Let us first calculate the self-energy of the ghost field.

%%%%%%%%%%%%%%%%%%%%%%%%%%%%%%%%%%%
\subsection{Self-energy of the Ghost Field}

Using the Feynman rules shown in Appendix A, the term that result after contracting the
indices and taking the trace of the structure functions is given by
\begin{eqnarray}
\widetilde{\Pi}(p)&=&\frac{im}{\kappa p^2}\half (c_f + c_d) \int \frac{d^3 k}{(2\pi)^3}
\frac{p^2k^2-(p\cdot k)^2}{k^2(k^2-m^2)(p+k)^2}\nonumber \\[1ex]
&&+\frac{im}{\kappa p^2}\half (c_f - c_d) \int \frac{d^3 k}{(2\pi)^3}
\frac{\cos(\widetilde{p}k)(p^2k^2-(p\cdot k)^2)}{k^2(k^2-m^2)(p+k)^2},\label{ghostsf}
\end{eqnarray}
where $\widetilde{p}k=p_\mu\Theta^{\mu\nu}k_\nu$ and $c_f$, $c_d$
are the quadratic Casimirs of the structure constants antisymmetric
and symmetric respectively. To obtain this factorization, in the process of using the Feynman rules we must take into account the properties of the algebra in the new $\Theta$-dependent structure function as is shown as follows
\begin{eqnarray}
\tr[F_{tsr}(p\wedge k)F_{usr}(p\wedge k)]&=& \tr \left\{\left[f_{tsr}
\cos\left(\frac{\widetilde{p}k}{2}\right)+d_{tsr}\sen\left(
\frac{\widetilde{p}k}{2}\right)\right]\right. \nonumber \\[1ex]
&&\left. \left[f_{usr}\cos\left(\frac{\widetilde{p}k}{2}\right)+d_{usr}
\sen\left(\frac{\widetilde{p}k}{2}\right)\right]\right\} \nonumber \\[1ex]
&=& \tr \left[ f_{tsr}\,f_{usr}\cos^2\left(\frac{\widetilde{p}k}{2}\right) +
d_{tsr}\,d_{usr}\sen^2\left(\frac{\widetilde{p}k}{2}\right)\right. \nonumber \\[1ex]
&&+\left. (f_{tsr}\,d_{usr} + f_{usr}\,d_{tsr})\cos\left(\frac{\widetilde{p}k}{2}\right)
\sen\left(\frac{\widetilde{p}k}{2}\right)\right],
\label{efeefenc}
\end{eqnarray}
which can be simplified by using the Jacobi identity $[T^k,\{T^l,T^m\}]+$
cyclic permutations $=0$. Thus we obtain
\eq
f_{klo}\,d_{mno}+f_{mlo}\,d_{nko}+f_{nlo}\,d_{kmo}=0.
\eqn
For our particular case we have
\begin{eqnarray}
f_{tsr}\,d_{usr}+f_{usr}\,d_{str}+f_{ssr}\,d_{tur}&=&0, \nonumber \\
f_{tsr}\,d_{usr}+f_{usr}\,d_{str}&=&0.
\label{efes}
\end{eqnarray}
After substitution of Eq. (\ref{efes}) in (\ref{efeefenc}) obtain finally that \eq
\tr[F_{tsr}(p\wedge k)F_{usr}(p\wedge
k)]=\tr[f_{tsr}\,f_{usr}]\cos^2\left(\frac{\widetilde{p}k}{2}\right)
+\tr[ d_{tsr}\,d_{usr}]\sen^2\left(\frac{\widetilde{p}k}{2}\right),
\eqn
where it is defined the quadratic Casimir as
\eq
\tr[f_{tsr}f_{usr}]=c_f \qquad\qquad \tr[ d_{tsr}d_{usr}]=c_d.
\eqn
Using some trigonometric properties we obtain the desired form (\ref{ghostsf}).

In Eq. (\ref{ghostsf})
we can see that the planar and non-planar contributions for this
diagram are separated. For computing the integrals we use the Feynman
parametrization
\eq
\frac{1}{abc}=\Gamma(3)\int_0^1 dx \int_0^{1-x}dy \frac{1}{[a(1-x-y)+bx+cy]^3},
\eqn
in which if we take $a=k^2$, $b=(k+p)^2$ and $c=k^2-m^2$ we get
\eq
\frac{1}{k^2(k^2-m^2)(k+p)^2}= 2\int_0^1 dx \int_0^{1-x}dy \frac{1}{[(k+xp)^2+x(1-x)p^2 -ym^2]^3}.
\eqn
Making the change of variable
\eq
k'=k+xp \qquad  \qquad M^2=ym^2-x(1-x)p^2,
\eqn
we can rewrite Eq. (\ref{ghostsf}) as the planar and non-planar contributions
\eq
\widetilde{\Pi}(p) = \widetilde{\Pi}_p(p) + \widetilde{\Pi}_{np}(p)
\eqn
where
\eq
\widetilde{\Pi}_p(p)=\frac{im}{\kappa p^2} (c_f + c_d)\int_0^1 dx \int_0^{1-x}dy
\int \frac{d^3 k'}{(2\pi)^3} \frac{p^2k'^2-(p\cdot k')^2}{[k'^2-M^2]^3}
\label{ppart}
\eqn
and
\eq
\widetilde{\Pi}_{np}(p) = \frac{im}{\kappa p^2} (c_f - c_d)\int_0^1 dx \int_0^{1-x}dy \int \frac{d^3 k'}{(2\pi)^3} \frac{\cos(\widetilde{p}k')(p^2k'^2-(p\cdot k')^2)}{[k^2-M^2]^3}.
\label{nppart}
\eqn
It is convenient to reduce this integral into a simpler form, for
which we use the property $\int d^Dk k^\mu k^\nu f(k^2)= \int d^D
k \,k^2f(k^2)\frac{\eta^{\mu\nu}}{D}$ and make
a Wick's rotation by taking  $k_0=ik_{E0}$, then $k_E^2=-k^2$ and $d^D k=id^Dk_E.$ Therefore we  write the planar part as\footnote{From now on we will omit the apostrophe in $k$ except that it does not cause confusion.}:
\eq
\widetilde{\Pi}_p(p)=-\frac{m}{\kappa p^2}\frac{2}{3} (c_f +
c_d)\int_0^1 dx \int_0^{1-x}dy \int \frac{d^3 k}{(2\pi)^3}
\frac{k^2}{[k^2+M^2]^3}.
\eqn
It is convenient use spherical coordinates such that $d^3k=d\Omega k^2 dk$. Integration over the
angles and using the definition of the beta function and its properties we find
\eq
\widetilde{\Pi}_p(p)=-\frac{m}{\kappa}\frac{1}{8}\frac{1}{2\pi} (c_f
+ c_d)\int_0^1 dx \int_0^{1-x}dy \frac{1}{[m^2y-x(1-x)p^2]^\half}.
\label{gh1}
\eqn
The non-planar part  (\ref{nppart}) after Wick's rotation is expressed as
\eq
\widetilde{\Pi}_{np}(p)=-\frac{m}{\kappa}\frac{2}{3} (c_f - c_d)\int_0^1 dx
\int_0^{1-x}dy \int \frac{d^3 k}{(2\pi)^3} \frac{k^2\cos(\widetilde{p}k)}{[k^2+M^2]^3}.
\label{ghnp}
\eqn
Defining a new variable $\sqrt{\widetilde{p}^2}k_\mu = z_\mu$  one can rewrite Eq. (\ref{ghnp}) as
\eq
\widetilde{\Pi}_{np}(p)=-\frac{m}{\kappa}\frac{2}{3}\frac{\rho}{(2\pi)^3} (c_f - c_d)
\int_0^1 dx \int_0^{1-x}dy \int d^3z \frac{z^2 \cos(z\cdot \widehat{p})}{[z^2+a^2]^3},
\label{nonperturbcontri}
\eqn
where we defined $a^2=M^2\rho^2$ and $\rho=\sqrt{\widetilde{p}^2}$. The last integral
in the previous expression can be rewritten as
\eq
I(a)= \frac{1}{8a^2}\left[\frac{d^2}{da^2}-\frac{1}{a}\frac{d}{da}\right]
\int d^3z \frac{z^2 \cos(z\cdot \widehat{p})}{z^2+a^2},
\label{aaaintegral}
\eqn
where $\widehat{p}$ is the unit vector along  $\widetilde{p}$. The integral arising in Eq. (\ref{aaaintegral}) can   be done by choosing, without loss of generality,  $z_2$ in the direction of $\widehat{p}$ \cite{yi}. For the integration we use the functional form of the modified Bessel function \cite{tabla}. Thus one finally gets for $I(a)$ the following form
\eq
I(a)=-\frac{2\pi^2}{8a^2}(a^2-3a)e^{-a}.
\eqn
Finally the non-planar correction of the ghost fields is
\eq
\widetilde{\Pi}_{np}(p)= \frac{m}{\kappa}\frac{2}{3}\frac{\rho}{(2\pi)^3} (c_f - c_d)
\int_0^1 dx \int_0^{1-x}dy \frac{2\pi^2}{8a^2}(a^2-3a)e^{-a},
\eqn
or in terms of $M$ we have
\eq
\widetilde{\Pi}_{np}(p)= \frac{m}{\kappa}\frac{\rho}{24}\frac{1}{2\pi} (c_f - c_d)
\int_0^1 dx \int_0^{1-x}dy \left(1-\frac{3}{\rho M}\right)e^{-\rho M}.
\eqn

%%%%%%%%%%%%%%%%%%%%%%%%%%%%%%%%%%%%%%%%
\subsection{Self-energy of the gauge field}

As was mentioned above, we are interested in noncommutative corrections to the
renormalization of the Chern-Simons coefficient, for this reason  in what follows we consider only the odd part of the self-energy of the gluon.

There are seven one-loop diagrams that contribute to the gauge field self-energy, but for the odd part $\Pi_o$, only the diagrams  that have a gluon loop and two of them that have a fermion loop do contribute (see Appendix B).  Due to supersymmetry the self-energy for the gauge field
have not UV divergencies and do not be necessary to regularize. Moreover, in \cite{asano,ferrari1,ferrari2,ferrari3,ferrari4} it was shown that the noncommutative supersymmetric Chern-Simons is indeed finite. The contribution to the
term $\Pi_o$ will be divided into a bosonic part $\Pi_o^B(p)$ for which only the ghost loop diagram contribute,
and a fermionic part $\Pi_o^F(p)$ where everything else contribute. In both parts there are planar and non-planar contributions. Thus we have for the bosonic part
\eq
\Pi_o^B(p) = \Pi_{op}^B(p) + \Pi_{onp}^B(p),
\eqn
where
\eq
\Pi_{op}^B(p)=\frac{im}{p^2}\half (c_f+c_d)\int \frac{d^3 k}{(2\pi)^3}
\frac{[k^2p^2-(k\cdot p)^2][5k^2+5k\cdot p+4p^2-2m^2]}{k^2(k^2-m^2)(k+p)^2[(k+p)^2-m^2]},
\label{unoplanar}
\eqn
and
\eq
\Pi_{onp}^B(p)= \frac{im}{p^2} \half (c_f-c_d)\int \frac{d^3 k}{(2\pi)^3}
\frac{\cos(\widetilde{p} k) [k^2p^2-(k\cdot p)^2][5k^2+5k\cdot p+4p^2-2m^2]}
{k^2(k^2-m^2)(k+p)^2[(k+p)^2-m^2]}.
\label{dosnoplanar}
\eqn
The fermionic contribution is given by
\eq
\Pi_o^F(p) = \Pi_{op}^F(p) + \Pi_{onp}^F(p),
\eqn
where
\eq
\Pi_{op}^F(p)=-\frac{im}{p^2}\half (c_f+c_d) \int \frac{d^3 k}{(2\pi)^3}
\frac{ 2p^2}{[(k+p)^2-m^2](k^2-m^2)},
\label{fodd1}
\eqn
and
\eq
\Pi_{onp}^F(p)=-\frac{im}{p^2}\half (c_f-c_d) \int  \frac{d^3 k}{(2\pi)^3}
\frac{\cos(\widetilde{p}k) 2p^2}{[(k+p)^2-m^2](k^2-m^2)}.
\label{fodd2}
\eqn
For simplicity we first calculate the planar part (\ref{fodd1}). Using the Feynman parametrization
$\frac{1}{ab}=\int_0^1 \frac{dx}{[ax+b(1-x)]^2}$ and
making $k'=k-xp$ and $M_1^2=m^2-x(1-x)p^2$, (\ref{fodd1}) is simplified
\eq
\Pi_{op}^F(p)=-im (c_f+c_d) \int_0^1dx\int \frac{d^3 k'}{(2\pi)^3} \frac{1}{[k'^2-M_1^2]^2}.
\eqn
Making a Wick's rotation and integrating in spherical coordinates this integral becomes
\eq
\Pi_{op}^F(p)=\frac{m}{4}\frac{1}{2\pi}\frac{1}{8} (c_f+c_d)\int_0^1 \frac{dx}{[m^2-x(1-x)p^2]^{1/2}}.
\eqn

The non-planar part, after Feynman parametrization and Wick's rotation, is written as
\eq
\Pi_{onp}^F(p)=m (c_f-c_d) \int_0^1dx\int \frac{d^3 k'}{(2\pi)^3}
\frac{\cos(\widetilde{p}k')}{[k'^2+M^2]^2},
\eqn
where we have defined $\sqrt{\widetilde{p}^2}k_\mu = z_\mu$ and
$a^2=M^2\rho^2$, like in the previous section,  we have
\eq
\Pi_{onp}^F(p)=m \frac{\rho}{(2\pi)^3}(c_f-c_d) \int_0^1dx
\int d^3 z\frac{\cos(\widehat{p}\cdot z)}{[z^2+a^2]^2}.
\eqn
The second integral in this equation reads
\eq
I_1(a)= -\frac{1}{2a}\frac{d}{da} \int d^3 z\frac{\cos(\widehat{p}\cdot z)}{z^2+a^2}.
\label{iuno}
\eqn
Following a similar procedure in the computation of the integral in the non-planar case for the ghost field (\ref{nonperturbcontri}) we obtain
\eq
I_1(a)= \frac{\pi^2 e^{-a}}{a}.
\eqn
Finally the non-planar contribution is given by
\eq
\Pi_{onp}^F(p)=\frac{m}{4} \frac{\rho}{2\pi}(c_f-c_d) \int_0^1dx \frac{e^{-\rho M_1}}{\rho M_1}.
\eqn

For the bosonic part the procedure is completely analogous though a bit more involved. Using the Feynman parametrization
\eq
\frac{1}{abcd}=3!\int_0^1dw\int_0^{1-x}dx \int_0^{1-w-x}dy \frac{1}{[ay+bx+cw+d(1-w-x-y)]^4},
\eqn
the planar part (\ref{unoplanar}) reads
\eq
\Pi_{op}^B(p)=3!\frac{2im}{3p^2}\half (c_f+c_d)\int_0^1dw\ dx\ dy\int \frac{d^3 k'}{(2\pi)^3}
\frac{k'^2p^2[5k'^2+p^2(5u^2-5u)+2m^2]}{[k'^2-M_2^2]^4},
\eqn
where $k'=k-up$,  $M_2^2=(w+y)m^2-u(1-u)p^2$ and $u=x+y$.

Making the Wick's rotation as in the previous cases and integrating out in spherical coordinates we obtain
\begin{eqnarray}
\Pi_{op}^B(p)&=&-\frac{2m}{3}\half (c_f+c_d)\int_0^1dw\ dx\ dy
\left[\frac{15}{16}\frac{1}{2\pi}\frac{5}{[(w+y)m^2-u(1-u)p^2]^{1/2}}\right.\nonumber \\[2ex]
&& \qquad\qquad\qquad\qquad -\left. \frac{3}{16}\frac{1}{2\pi}\frac{p^2(5u^2-5u)-2m^2}{[(w+y)m^2-u(1-u)p^2]^{3/2}}\right].
\end{eqnarray}

The non-planar part after parametrization and Wick's rotation is given by
\begin{eqnarray}
\Pi_{onp}^B(p)&=&-3!\frac{2m}{3}\half (c_f-c_d)\int_0^1dw\ dx\ dy\left\{ \int \frac{d^3 k}{(2\pi)^3}
\frac{5k^4\cos(\widetilde{p}k)}{[k^2+M_2^2]^4}\right. \nonumber \\[2ex]
&&-\left. [p^2(5u^2-5u)-2m^2]\int \frac{d^3 k}{(2\pi)^3}
\frac{\cos(\widetilde{p}k)k^2}{[k^2+M_2^2]^4}\right\}.
\end{eqnarray}
We make the same change of variables that in the previous cases and we get
\begin{eqnarray}
\Pi_{onp}^B(p)&=&-3!\frac{2m}{3}\frac{5\rho}{(2\pi)^3}\half (c_f-c_d)\int_0^1dw\ dx\ dy\left\{ \int d^3 z
\frac{z^4\cos(\widehat{p}\cdot z)}{[z^2+a^2]^4}\right.\nonumber \\[2ex]
&&-\left.\rho^2 [p^2(5u^2-5u)-2m^2] \int d^3 z
\frac{\cos(\widehat{p}\cdot z)z^2}{[z^2+a^2]^4}\right\}.
\end{eqnarray}
The last integrals in each term can be written as
\eq
I_2(a)=-\frac{1}{48a^3}\left[\frac{d^3}{da^3}-\frac{3}{a}\frac{d^2}{da^2}+\frac{3}{a^2}\frac{d}{da}\right]\int d^3 z \frac{z^4\cos(\widehat{p}\cdot z)}{z^2+a^2},
\label{idos}
\eqn
\eq
I_3(a)=-\frac{1}{48a^3}\left[\frac{d^3}{da^3}-\frac{3}{a}\frac{d^2}{da^2}+\frac{3}{a^2}\frac{d}{da}\right]\int d^3 z \frac{\cos(\widehat{p}\cdot z)z^2}{z^2+a^2}
\label{itres}.
\eqn
Similarly than the previous situations we can compute these integrals and this yields
\eq
I_2(a)=\frac{2\pi^2}{48a^3}(a^4+15a^3+15a^2)e^{-a},
\eqn
\eq
I_3(a)=\frac{2\pi^2}{48a^3}(a^2-3a-3)e^{-a}.
\eqn
Finally we obtain that the correction is given by
\begin{eqnarray}
\Pi_{onp}^B(p)&=& -\frac{5m\rho}{48}\frac{1}{2\pi}(c_f-c_d)\int_0^1dw\ dx\ dy\left\{ \left(\rho M_2+15+\frac{15}{\rho M_2}\right)e^{-\rho M_2} \right. \nonumber \\
&& -\left. [p^2(5u^2-5u)-2m^2]\left(\frac{1}{\rho M_2}-\frac{3}{(\rho M_2)^2}-\frac{3}{(\rho M_2)^3}\right)e^{-\rho M_2}\right\}.
\end{eqnarray}

%%%%%%%%%%%%%%%%%%%%%%%%%%%%%%%%%%%%%%%%%
%%%%%%%%%%%%%%%%%%%%%%%%%%%%%%%%%%%%%%%%%
%%%%%%%%%%%%%%%%%%%%%%%%%%%%%%%%%%%%%%%%%
\section{Shift of $\kappa$}
In order to calculate the shift of the $\kappa$ coefficient we will expand the contributions
to the self-energy of gluon, and the contribution of the ghost fields and integrate over Feynman parameters
obtaining in this way that
\begin{eqnarray}
\widetilde{\Pi}(p)&\approx& -\frac{c_f}{6\pi|\kappa|}+\frac{m\rho(c_f-c_d)}{24\pi},
\\
\Pi_o^F(p)&\approx&\frac{c_f}{4\pi|\kappa|}-\frac{m\rho(c_f-c_d)}{8\pi},
\\
\Pi_o^B(p)&\approx& -\frac{7c_f}{12\pi|\kappa|}.
\end{eqnarray}

We can see that for the bosonic part of the self-energy that comes from the gluon there is not
correction due to noncommutativity. The value obtained is the same that the obtained for the
commutative case with $p=0$. The terms that have not as common factor $\rho$
in the fermionic and ghost contributions are precisely those that correspond to the
commutative usual case. The other terms are due to the non-commutativity.

Finally applying the equation
\eq
\kappa_{ren}=\kappa\left(1-\frac{1}{k}\Pi_o(p)+2\widetilde{\Pi}(p)\right),
\eqn
we obtain the result
\begin{equation}
\kappa_{ren}=\kappa \left(1+\frac{5}{24\pi}g^2 \Theta (c_f-c_d)p\right).
\label{correccionuno}
\end{equation}

For finding the shifts for the ${\cal N}=2$ theory it is necessary to consider that
$C_1=C_2=\lambda_3=\chi=0$, but as in the fermionic contribution to the
ghost self-energy, the contribution of $\lambda_a$ is canceled by the
contribution of $\chi$. Then we have that for the ${\cal N}=2$ theory the shift is the same.
Nevertheless for the ${\cal N}=1$ theory we can obtain the contributions from
the ${\cal N}=2$ theory by considering that  $C_3=0$ and  $\lambda_2=0$ so the contribution to
the fermionic part of the self-energy is one-half of the result presented here.
In this way we find that for the ${\cal N}=1$ theory we have
\eq
\kappa_{ren}=\kappa \left(1+\frac{c_f}{8\pi|\kappa|}+\frac{5}{48\pi}g^2 \Theta (c_f-c_d)p\right).
\label{correcciondos}
\eqn

%%%%%%%%%%%%%%%%%%%%%%%%%%%%%%%
%%%%%%%%%%%%%%%%%%%%%%%%%%%%%%%
%%%%%%%%%%%%%%%%%%%%%%%%%%%%%%%
\section{Final Comments}

In the present paper a noncommutative version of the supersymmetric YM-CS theory is studied. This theory constitutes a Moyal deformation of the theory considered in \cite{kao}. For this noncommutative deformation we calculated the shifting to the Chern-Simons coefficient due to noncommutativity in the limit of small moments. This calculation was done in the context of perturbative ${\cal N}=1,2,3$ supersymmetric YM-CS gauge theory in three dimensions with compact gauge group $U(N)$. It was found that this shift have a dependence of noncommutative parameter $\Theta$ and the momenta $p$ (see Eqs. (\ref{correccionuno})  and (\ref{correcciondos})). This correction, nevertheless vanishes in the limit $\Theta \to 0$ which is expected.

Although we explore noncommutative gauge theories in the perturbative context it is known that the analyticity properties of the obsevables of the theory with respect to the noncommutative parameter has information about non-perturbative properties of the system \cite{AlvarezGaume:2001tv} and there were computed different nonperturbative quantities as the Witten index \cite{Witten:1999ds}. It is known that that Witten's index is compatible with a one-loop quantum correction to the Chern-Simons coupling $\kappa$ in the Yang-Mills-Chern-Simons gauge theory. Given our result from Eqs. (\ref{correccionuno})  and (\ref{correcciondos})  it would be very interesting to explore if there will be a modification introduced by the noncommutative theory and make a comparison with the result in \cite{AlvarezGaume:2001tv}.  We are currently exploring these issues and intend to report some progress elsewhere.

\centerline{\bf Acknowledgments}

The work of S.E.-J. was supported by a CONACyT postdoctoral
fellowship and Grant PROMEP /103.5/08/3291. The work of H.G.-C.
was supported in part by CONACyT M\'exico Grant 128761.

%%%%%%%%%%%%%%%%%%%%%%%%%%%%%%%%%%%%%%%%%%%%%%%%%%%%%%%%%%%%%%%%%%%%%%%%%
%%%%%%%%%%%%%%%%%%%%%%%%%%%%%%%%%%%%%%%%%%%%%%%%%%%%%%%%%%%%%%%%%%%%%%%%%
%%%%%%%%%%%%%%%%%%%%%%%%%%%%%%%%%%%%%%%%%%%%%%%%%%%%%%%%%%%%%%%%%%%%%%%%%
\newpage
\appendix
\section{Feynman rules}
The propagator for the gauge bosons  $A_\mu$ is
\eq
\parbox{30mm}{\propagluon}
\Delta_{\mu\nu}(k)=\frac{g^2}{k^2(k^2-m^2)}(k_\mu k_\nu-k^2\eta_{\mu\nu}-
im\varepsilon_{\mu\nu\rho}k^\rho)+g^2\xi\frac{k_\mu k_\nu}{k^4},
\eqn
the propagator for each $\lambda_a$ is given by
\eq
\parbox{3cm}{\propalambda}
D^{mn}(k)=\frac{\delta^{mn}g^2}{\sh k-m},\qquad\qquad\qquad\qquad\qquad\qquad\qquad
\eqn
the propagator for the fermion $\chi$ is
\eq
\parbox{3cm}{\propachi}
{\cal D}^{mn}(k)=\frac{\delta^{mn}g^2}{\sh k+m},\qquad\qquad\qquad\qquad\qquad\qquad\qquad
\eqn
the propagator for the bosons $C_a$ is\\[1ex]
\eq
\parbox{3cm}{\propaboson}
\delta(k)=-\frac{g^2}{ k^2+m^2}, \qquad\qquad\qquad\qquad\qquad\qquad\qquad
\eqn
the propagator for the ghost fields is $\eta$ son\\[1ex]
\eq
\parbox{3cm}{\propaghost}
\widetilde{\Delta}_{ab}(k)=-\frac{\delta_{ab}g^2}{k^2}.\qquad\qquad\qquad\qquad\qquad\qquad\qquad
\eqn
\\[2ex]
The Feynman rules for the vertex are:\\[2ex]
\eq
\parbox{30mm}{\tresgvertex}
 =\frac{-i}{g^2}[(p-r)^\nu\eta^{\rho\mu}+(r-q)^\mu\eta^{\nu\rho}+(q-p)^\rho\eta^{\mu\nu}
-im\varepsilon^{\mu\nu\rho}]F_{klm}(q\wedge p),
\eqn
\\[2ex]
\eq
\parbox{30mm}{\fourgvertex}
=\begin{array}{rl}
 &\frac{1}{g^2}\bigg[ F_{nmc}(r\wedge s)F_{lkc}(p\wedge q)(\eta^{\mu\rho}\eta^{\nu\delta}-
\eta^{\mu\delta}\eta^{\nu\rho}) \\
&+ F_{nlc}(q\wedge s)F_{mkc}(p\wedge r)(\eta^{\mu\nu}\eta^{\rho\delta}-
\eta^{\mu\delta}\eta^{\nu\rho})\\
&+ F_{nkc}(p\wedge s)F_{mlc}(q\wedge r)(\eta^{\mu\nu}\eta^{\rho\delta}-
\eta^{\mu\rho}\eta^{\nu\delta})\bigg],\quad
\end{array}
\eqn
\\[2ex]
\eq
\parbox{30mm}{\lamgvertex}
 =\frac{-i}{2g^2},\gamma^\mu F_{klm}(q\wedge p)\delta_{ab},\qquad\qquad\qquad\qquad\qquad\qquad\qquad
\eqn
\\[2ex]
\eq
\parbox{30mm}{\chigvertex}
 =\frac{-i}{2g^2},\gamma^\mu F_{klm}(q\wedge p),\qquad\qquad\qquad\qquad\qquad\qquad\qquad
\eqn
\\[2ex]
\eq
\parbox{30mm}{\ghgvertex}
 =-ir^\mu F_{klm}(q\wedge p),\qquad\qquad\qquad\qquad\qquad\qquad\qquad\qquad
\eqn
\\[2ex]
\eq
\parbox{30mm}{\tsgvertex}
 =\frac{1}{g^2}ir^\mu F_{klm}(q\wedge p)\delta_{ab},\qquad\qquad\qquad\qquad\qquad\qquad\qquad\qquad
\eqn
\\[2ex]
\eq
\parbox{30mm}{\fsgvertex}
 =\frac{1}{g^2}\delta^{\mu\nu}\delta_{ab} [F_{lnt}(s\wedge q)F_{kmt}(r\wedge p)  +
F_{knt}(s\wedge p)F_{lmt}(r\wedge q)],
\eqn
\\[2ex]
\eq
\parbox{30mm}{\slamvertex}
 =\half\frac{1}{g^2} \varepsilon_{cba} F_{klm}(q\wedge p),\qquad\qquad\qquad\qquad\qquad\qquad\qquad\qquad
\eqn
\\[2ex]
\eq
\parbox{30mm}{\slamchivertex}
 =\frac{1}{g^2} F_{klm}(q\wedge p)\delta_{ab},\qquad\qquad\qquad\qquad\qquad\qquad\qquad\qquad
\eqn
\\[2ex]
\eq
\parbox{30mm}{\tscalvertex}
 =\frac{1}{g^2}\varepsilon_{abc} F_{klm}(p\wedge q),\qquad\qquad\qquad\qquad\qquad\qquad\qquad\qquad
\eqn
\\[2ex]
\eq
\parbox{30mm}{\fscalvertex}
=\begin{array}{rl}
 &\frac{1}{g^2}\bigg[ F_{nmc}(r\wedge s)F_{lkc}(p\wedge q)(\delta^{ac}\delta^{bd}-\delta^{ad}\delta^{bc}) \\
&+ F_{nlc}(q\wedge s)F_{mkc}(p\wedge r)(\delta^{ab}\delta^{cd}-\delta^{ad}\delta^{bc})\\
&+ F_{nkc}(p\wedge s)F_{mlc}(q\wedge r)(\delta^{ab}\delta^{cd}-\delta^{ac}\delta^{bd})\bigg],
\quad\qquad\qquad
\end{array}
\eqn
\\[2ex]
where the indexes $a,b,c,d$ run from 1 to 3 and refer to $C_a$ y $\lambda_a$ and $k,l,m,n$ refer to the group algebra indexes.

\section{One loop diagrams}
\begin{tabular}{ccc}
\parbox{50mm}{\loopuno}& \parbox{50mm}{\loopdos} &\parbox{50mm}{\looptres} \\[4ex]
(a)&(b)&(c)\\[4ex]
\parbox{50mm}{\loopfour}& \parbox{50mm}{\loopfive} & \\[2ex]
(d)& (e) & \\[7ex]
\parbox{50mm}{\loopsix}&\parbox{50mm}{\loopseven}&\\[2ex]
(f)&(g)&
\end{tabular}

\end{fmffile}

\end{document}